\documentclass[article,12pt]{article}%

\usepackage{comment}
\usepackage{makecell}
\usepackage{tikz}
\usepackage{tabularx}
\usepackage{colortbl}
\usepackage{gensymb}
\usepackage{graphicx}
\usepackage{caption}
\usepackage{booktabs}
\graphicspath{{Fig/}}
\usepackage[hidelinks]{hyperref}
\usepackage{fbox}
\usepackage[margin=1in]{geometry}

\title{Critical review of conformational\\ B-cell epitope prediction methods}
\author{Gabriel Cia $^{1,2}$, Fabrizio Pucci $^{1,2,\dagger}$, Marianne Rooman $^{1,2,\dagger}$ \\  \\
        \small $^{1}$ Computational Biology and Bioinformatics, Universit\'{e} Libre de Bruxelles,\\\small  F. Roosevelt Avenue, 1050, Brussels, Belgium\\
        \small $^{2}$ Interuniversity Institute of Bioinformatics in Brussels, Belgium\\
        \small $^{\dagger}$Contributed equally to this work \\ \small Email: Gabriel.Cia@ulb.be, Fabrizio.Pucci@ulb.be, Marianne.Rooman@ulb.be
}
\date{} 

\begin{document}
\maketitle




\begin{abstract}
Accurate \textit{in-silico} prediction of conformational B-cell epitopes would lead to major improvements in disease diagnostics, drug design and vaccine development. A variety of computational methods, mainly based on machine learning approaches, have been developed in the last decades to tackle this challenging problem. Here, we rigorously benchmarked nine state-of-the-art conformational B-cell epitope prediction webservers, including generic and antibody-specific  methods, on a dataset of over 250 antibody-antigen structures. The results of our assessment and statistical analyses show that all the methods achieve very low performances, and some do not perform better than randomly generated patches of surface residues. In addition, we also found that commonly used consensus strategies that combine the results from multiple webservers are at best only marginally better than random. Finally, we applied all the predictors to the SARS-CoV-2 spike protein as an independent case study, and  showed that they  perform poorly  in general, which largely recapitulates our benchmarking conclusions. 
We hope that these results will \textcolor{black}{lead to greater caution when using these tools until the biases and issues that limit current methods have been addressed}, promote the use of state-of-the-art evaluation methodologies in future publications, and suggest new strategies to improve the performance of conformational B-cell epitope prediction methods.
\end{abstract}



\pagebreak

\section{Introduction}


The ever increasing amounts of biological data that are being generated and deposited in publicly accessible databases \cite{barrett2012ncbi, chen2017protein} have boosted the development of machine learning (ML) models that are being used to help in advancing a variety of  problems in the fields of genomics, proteomics and molecular evolution   \cite{larranaga2006machine, min2017deep}. The availability of 3-dimensional (3D) structural information from either experiments \cite{berman2000protein} or accurate prediction tools \cite{waterhouse2018swiss,jumper2021highly} has further led to substantial improvements of \textit{in-silico} prediction and modeling tools. One of the fields that has seen the development of a large number of structure-based ML models is B-cell epitope prediction. B-cell epitopes are typically protein surface regions which are bound by antibodies, and knowledge of the residues that form an epitope is key for unraveling disease mechanisms \cite{davidson2001autoimmune, burger2013b} or for applications such as vaccine design,  immunotherapy and immunoassay development \cite{ponomarenko2009b}. 

Several experimental methods are available to determine B-cell epitopes \cite{ponomarenko2009b}, but they  are expensive, time-consuming and some require a high level of lab expertise. This is why the development of \textit{in-silico} tools has attracted a lot of attention. Initial methods focused on linear B-cell epitopes and relied on features derived from antigen sequences, but early on their predictive power was shown to be no better than random \cite{blythe2005benchmarking}, a conclusion that was further confirmed in a recent study \cite{galanis2021linear}. As more X-ray structures of antibody-antigen complexes were deposited in the Protein Data Bank (PDB) \cite{berman2000protein}, a number of structure-based methods were developed to predict conformational or discontinuous B-cell epitopes, which contain residues that are not necessarily contiguous along the protein sequence. Many of these methods have reported significantly better than random predictive power \cite{greenbaum2007towards} (see \cite{sanchez2017fundamentals} for a historic presentation of B-cell epitope prediction methods).

Nevertheless, a number of voices \cite{sela2015antibody, van2019specificity, jespersen2019antibody, sela2015pease, sela2013structural, zhao2011antibody} have raised concerns regarding the feasibility of generic epitope predictions, \textit{i.e.} predicting all the  epitopes on a given antigen for all possible antibodies. Indeed, some evidence suggests that antibodies may be raised against virtually any part of the surface of any given protein \cite{benjamin1984antigenic, kringelum2013structural, kunik2013indistinguishability}, \textcolor{black}{except in the case of chemical modifications such as glycosylation which are known to often block antibody binding} \cite{chang2019glycosylation, casalino2020beyond, wintjens2020impact}. The case of extensively studied proteins such as lysozyme, HIV-gp120 and, more recently, the receptor binding domain (RBD) of the SARS-CoV-2 spike protein  show that it is possible to find epitopes on almost the entire surface of an antigen. If this were to be the general case, generic epitope prediction approaches would be futile. 

As a result, a new trend has emerged in the field that challenges the generic epitope prediction paradigm and instead attempts to develop antibody-specific epitope predictors
\cite{sela2015antibody}. The main advantage of this approach is that it deals with a more constrained and tractable task as opposed to generic epitope prediction. Its downside is that it requires prior knowledge of the antibodies that need to be screened, which greatly limits the number of use cases compared to generic epitope prediction methods. Moreover, current antibody-specific epitope predictors are not fast enough to screen even a small fraction of the space of all possible antibodies, which is currently estimated at 10\textsuperscript{12} for naive antibodies and up to 10\textsuperscript{16} - 10\textsuperscript{18} for all possible antibodies \cite{briney2019commonality}.

To advance the field of B-cell epitope prediction, we have benchmarked and analyzed some of the most popular generic and antibody-specific B-cell epitope prediction methods by testing whether they are able to accurately identify experimentally validated epitopes. This evaluation was performed on a dataset of over 250 non-redundant antibody-antigen structures using a rigorous benchmarking methodology.

\section{Materials and methods}

\subsection{Surface residues}

Residues were considered as part of the surface if they have a relative solvent accessible surface area (RSA) of at least 10\%. The RSA of a residue X in a given protein structure, expressed in \%, is defined as the sum of the accessible surface areas (ASA) of  its heavy atoms  
divided by its maximal ASA reached when included in a Gly-X-Gly tripeptide in extended conformation. The ASA and RSA values were computed using an in-house program \cite{dalkas2014cation}.

\subsection{Epitope residues}
Antigen surface residues residues that undergo a change in RSA of at least 5\% upon binding with an antibody ($\Delta RSA= RSA_{unbound} - RSA_{bound} \geq 5 \%$) were considered as epitope residues. 


\subsection{Structure datasets}

The structures of complete antibodies (heavy and light chain) in complex with protein antigens were downloaded from the AntiBody DataBase \cite{ferdous2018abdb} in PDB format. This represented 3,000 complexes at the date of 10/2020. Structures with a resolution greater than 3.0 {\AA}, an R-factor greater than  0.30, or in which less than 80\% of the residues have atomic coordinates were  overlooked. Complexes in which the antigen has less than 50 residues were also dropped. This resulted in a quality filtered set of 1,151 antibody-antigen structures. The epitopes on the antigens were determined as described in the previous subsection. This set 
is referred to as $\mathcal{E}_{Ag}$.

In order to avoid redundancy and correctly evaluate the benchmarked generic B-cell epitope predictors, the antigens from the $\mathcal{E}_{Ag}$ set were clustered according to their sequence identity using CD-hit \cite{li2006cd} with a 70\% sequence identity threshold. This yielded 268 distinct antigen clusters. The representative antigen structure of each cluster was chosen to be the one identified by CD-hit. The epitope residues of all antigens in a given cluster were mapped onto the representative antigen structure by aligning their sequences using Biopython's local alignment algorithm \cite{cock2009biopython} with the same default parameter settings as EMBOSS \cite{madeira2022search} (substitution matrix = BLOSUM62, open gap penalty = -10, extension gap penalty = -0.5); epitope residues were only mapped if the aligned residues were identical. The dataset of representative antigen structures with all epitopes mapped onto them is referred to as $\mathcal{E}_{Ag}^{rep}$.

\textcolor{black}{The list of structures of the two datasets $\mathcal{E}_{Ag}$ and $\mathcal{E}_{Ag}^{rep}$ along with their PDB files are available at \url{https://github.com/3BioCompBio/BCellEpitope}.}

\subsection{Evaluation metrics}

To estimate the prediction performance of the benchmarked predictors, \textcolor{black}{we used a number of well established performance metrics \cite{jiao2016performance}, including} the balanced accuracy (BAC), the Matthews correlation coefficient (MCC), the  area under the receiver operating characteristic curve (ROC-AUC) and the  area under the precision-recall curve (PR-AUC), defined as:
\begin{itemize}
 \setlength\itemsep{1.25em}

 \item \begin{math}
  \displaystyle
  \mathrm{
  BAC = \frac{1}{2} \left( \frac{TP}{TP+FN} + \frac{TN}{TN+FP} \right)
  }
 \end{math}
 
 \item \begin{math}
  \displaystyle
  \mathrm{
  MCC = \frac{TP\;TN - FP\;FN}{\sqrt{(TP+FP)(TP+FN)(TN+FP)(TN+FN)}}
  }
 \end{math}
 
 \item  ROC-AUC, \textit{i.e.} the area under the curve (AUC) of the recall or sensitivity (TP/(TP+FN)) versus the false positive rate or specificity (FP/(FP+TN)).
 
 \item  PR-AUC, \textit{i.e.} the AUC of the positive predictive value (PPV) or precision (TP/(TP+FP)) versus the recall or sensitivity (TP/(TP+FN)).
 
\end{itemize}

\noindent where TP are correctly predicted epitope residues, FP  non-epitope residues incorrectly predicted as epitope residues, TN  correctly predicted non-epitope residues, and FN  epitope residues incorrectly predicted as non-epitope residues. The mean random value is  equal to 0.5 for BAC and ROC-AUC, and 0 for MCC; for PR-AUC it is dataset-dependent.

\subsection{Random epitope prediction procedure}

In order to assess the statistical significance of the different methods against random predictions, we defined two  procedures, one that predicts random surface residues, and a second that predicts random patches of surface residues, \textit{i.e} groups of residues that are nearby in the 3D structure. Each procedure is repeated 2,000 times in order to generate bootstrap distributions that are then used to calculate p-values.

The first procedure randomly predicts $N_r$ random surface residues as epitopes on each antigen structure.  We tested two strategies for setting the value of $N_r$: (1) $N_r$ = 18, which corresponds to the average epitope size in our $\mathcal{E}_{Ag}$ dataset; (2)  $N_r$ chosen dynamically to match the number of residues predicted by each method for each structure. The latter strategy allowed us to assess how our random procedure compares to each method in the case of equivalent prediction thresholds, and led to method-specific bootstrap distributions for each metric (MCC, BAC, ROC-AUC, PR-AUC).

The above  procedure is overly simplistic as epitope residues are  not randomly scattered across the protein surface, but rather form patches of nearby surface residues. We therefore developed a second procedure that predicts $N_p$ random patches of $N_r$ surface residues each. The patches were constructed by randomly selecting a surface residue and adding its $N_r$ - 1 closest surface residues. Here we also used two strategies to set the values of $N_p$ and $N_r$: (1) $N_p=1$ and $N_r=18$;  (2) dynamical number of epitope residues $N$ matching the number of predicted residues on a per-method and per-structure basis, distributed in $N_p$ patches of $N_r$ residues as: $N_p=\lfloor N/18 \rfloor$ and $N_r=18$ for all  patches but  one for which $N_r=N-(N_p-1)*18$. 

\begin{table}[h]
\centering
\renewcommand{\arraystretch}{1.25} 
\begin{tabular}{lcccccc}
\hline
Method & ROC-AUC  & BAC & MCC & PR-AUC &  $N_{antigens}$ & $F_{predicted}$\\
\hline
\multicolumn{6}{l}{\textbf{Sequence-based generic methods}} \\
\hspace*{0.25em} BepiPred2 & 0.53  & 0.52 & 0.02 & 0.24 & 101/268 & 52 \% \\
\hspace*{0.25em} CBTOPE & 0.46  & 0.47 & -0.06 & 0.19 & 229/268 & 38 \% \\
\multicolumn{6}{l}{\textbf{Structure-based generic methods}} \\
\hspace*{0.25em} SEPPA3 & 0.53  & 0.51 & 0.00 & 0.23 & 105/268 & 47 \% \\
\hspace*{0.25em} DiscoTope2 & \textbf{0.58}  & \textbf{0.53} & \textbf{0.06} & 0.26 & 220/268 & 19 \% \\
\hspace*{0.25em} ElliPro & 0.56 & \textbf{0.53} & 0.04 & 0.23 & 259/268 & 63 \% \\
\hspace*{0.25em} EPSVR & 0.53  & 0.52 & 0.03 & 0.26 & 236/268 & 52 \% \\
\hspace*{0.25em} BEpro & \textbf{0.58}  & \textbf{0.53} & \textbf{0.06} & 0.27 & 235/268 & 12 \% \\
\hspace*{0.25em} epitope3D & 0.41  & 0.41 & -0.17 & 0.09 & 221/268 & 3 \% \\
\multicolumn{6}{l}{\textbf{Structure-based antibody-specific methods}} \\
\hspace*{0.25em} EpiPred & 0.50  & 0.50 & -0.01 & \textbf{0.35} & 746/1151 & 49 \% \\
\arrayrulecolor{black}\hline
\end{tabular}

\vspace{0.25cm}

\caption{Average performance of each of the benchmarked conformational B-cell epitope predictors. The highest score(s) of each metric are in bold. $N_{antigens}$ corresponds to the number of structures on which each method has been evaluated out of the total number of eligible antigens. $F_{predicted}$ is the mean fraction (in \%) of surface residues predicted as epitopes. \textcolor{black}{Note that the reason why DiscoTope2 and BEpro have a low $F_{predicted}$ comes from the fact that these methods use very high prediction thresholds.}}
\label{table:benchmark_metrics}
\end{table}

\section{Results and discussion}

\subsection{Epitope dataset analysis}

We used two datasets: $\mathcal{E}_{Ag}$ that contains  1,151 good-quality antigen structures, each carrying a single epitope, and the non-redundant and non-homologous $\mathcal{E}_{Ag}^{rep}$ dataset,  which contains 268 representative antigen structures onto which we mapped all the  epitopes identified on homologous structures in $\mathcal{E}_{Ag}$ (see Methods for details). Mapping multiple known epitopes onto a single antigen structure prevents as much as possible erroneous false positive annotations that would arise if different epitopes from the same antigen were evaluated independently from each other \cite{kringelum2012reliable}.

The number of mapped epitopes per antigen structure in $\mathcal{E}_{Ag}^{rep}$ follows a decreasing exponential-type distribution in which 85\% of the structures have less than 5 mapped structures (see \textcolor{black}{Supplementary Figure S1}). For some extensively studied antigens such as lysozyme and HIV-gp-120, this number increases to more than 35. In the case of lysozyme, the epitopes cover almost the entire surface: 70 out of 85 surface residues belong to at least one of the many lysozyme epitopes found in our $\mathcal{E}_{Ag}$ dataset.
The generality of this observation is currently an open question; we will come back to it in the discussion section.

\subsection{Benchmarking methodology}

We assessed conformational B-cell epitope prediction methods with a functioning webserver. The list of methods that were selected includes two generic sequence-based methods: Bepipred-2.0 \cite{jespersen2017bepipred} and CBTOPE \cite{ansari2010identification}; six generic structure-based methods: SEPPA3 \cite{zhou2019seppa}, DiscoTope2 \cite{kringelum2012reliable}, ElliPro \cite{ponomarenko2008ellipro}, EPSVR \cite{liang2010epsvr}, BEpro \cite{sweredoski2008pepito} and epitope3D
\cite{da2022epitope3d}; and one antibody-specific structure-based predictor: EpiPred \cite{krawczyk2014improving}.

For evaluating each of the generic epitope prediction tools, we used the subset of the $\mathcal{E}_{Ag}^{rep}$ dataset that is not contained in the training dataset of the method considered. More precisely, we removed from $\mathcal{E}_{Ag}^{rep}$ any antigen that has a sequence identity of more than 99\% with any antigen in the training dataset of the  method that is being assessed. \textcolor{black}{Note that using a lower sequence identity threshold of 70\% has virtually no effect  on the score values reported in \autoref{table:benchmark_metrics}, as seen in Table S1.} For assessing the antibody-specific epitope prediction tool EpiPred, we used  the $\mathcal{E}_{Ag}$ set from which we removed all the PDB structures that are included in the method's training set. 
Although this procedure \textcolor{black}{means} that all the methods were assessed on different test sets, \textcolor{black}{it avoids biases due to evaluating training data.} 

Furthermore, we only considered the predictions made for surface residues (as defined in Methods) in the assessment, as core residues can not be part of B-cell epitopes. Note that the prediction scores would be much better if both surface and core residues were considered. This does not make sense for structure-based predictors and basically boosts sequence-based predictor performance given that the identification of surface residues is an easier problem than epitope prediction. 

We used the threshold-independent metric MCC and BAC for predictors evaluation as they account for all the categories of the confusion matrix.
In addition, we also used  ROC-AUC and PR-AUC as these performance metrics are independent of any classification threshold value and thus give complementary information. 

\subsection{Benchmarking results}

We report the average BAC, MCC, ROC-AUC and PR-AUC scores of the benchmarked methods in \autoref{table:benchmark_metrics} and their statistical significance against different random procedures in \textcolor{black}{Table S2a}. \textcolor{black}{Additional metrics for evaluating the methods, including sensitivity, specificity, precision and F1 score, are available in Table S3}. What clearly comes out is that all the methods have very low performances, as indicated by ROC-AUC and BAC values $< 0.6$  and MCC values $< 0.1$. BEpro and DiscoTope2 are the highest scoring methods, both achieving identical metrics (ROC-AUC=0.58, BAC=0.53, MCC=0.06). In contrast, the scores of epitope3D  are even worse than random (ROC-AUC=0.41, BAC=0.41, MCC=-0.17), because almost all  its predicted epitope residues are situated in the protein core. The first conclusion we can draw from these results is that even the highest scoring methods have very little predictive power. 

Surprisingly, the antibody-specific epitope predictor, EpiPred, does not show better overall performance than the best generic epitope predictors in terms of ROC-AUC, BAC and MCC. However, it does have the highest PR-AUC score, which indicates that the knowledge of the antibody improves the precision of the predicted epitopes. 

\textcolor{black}{Note that all  these results are independent of the chosen definition of epitopes. Indeed, comparing \autoref{table:benchmark_metrics} with Table S4, we observe that the scores are almost identical whether using an RSA- or distance-based definition of epitope residues, the latter being another common definition used by many of the benchmarked methods.
Moreover, we  analyzed how the definition of surface residues influences the methods' scores. For that purpose, we computed the MCC scores as a function of the RSA thresholds used for defining surface residues, as shown in Figure S2. 
We see that all the prediction methods have systematically worse scores as the threshold increases, which indicates that the more we restrict the evaluation to residues that are truly at the surface, the worse the methods perform.
Conversely, considering buried residues as belonging to the surface makes the predictions easier, because it artificially increases the difference between epitopes and non-epitopes  by basically enriching the latter  with hydrophobic residues much more than the former that are defined by an additional threshold on $\Delta {RSA}$  (see Methods). }

\begin{figure*}[t]
\centering
\includegraphics[width=\linewidth]{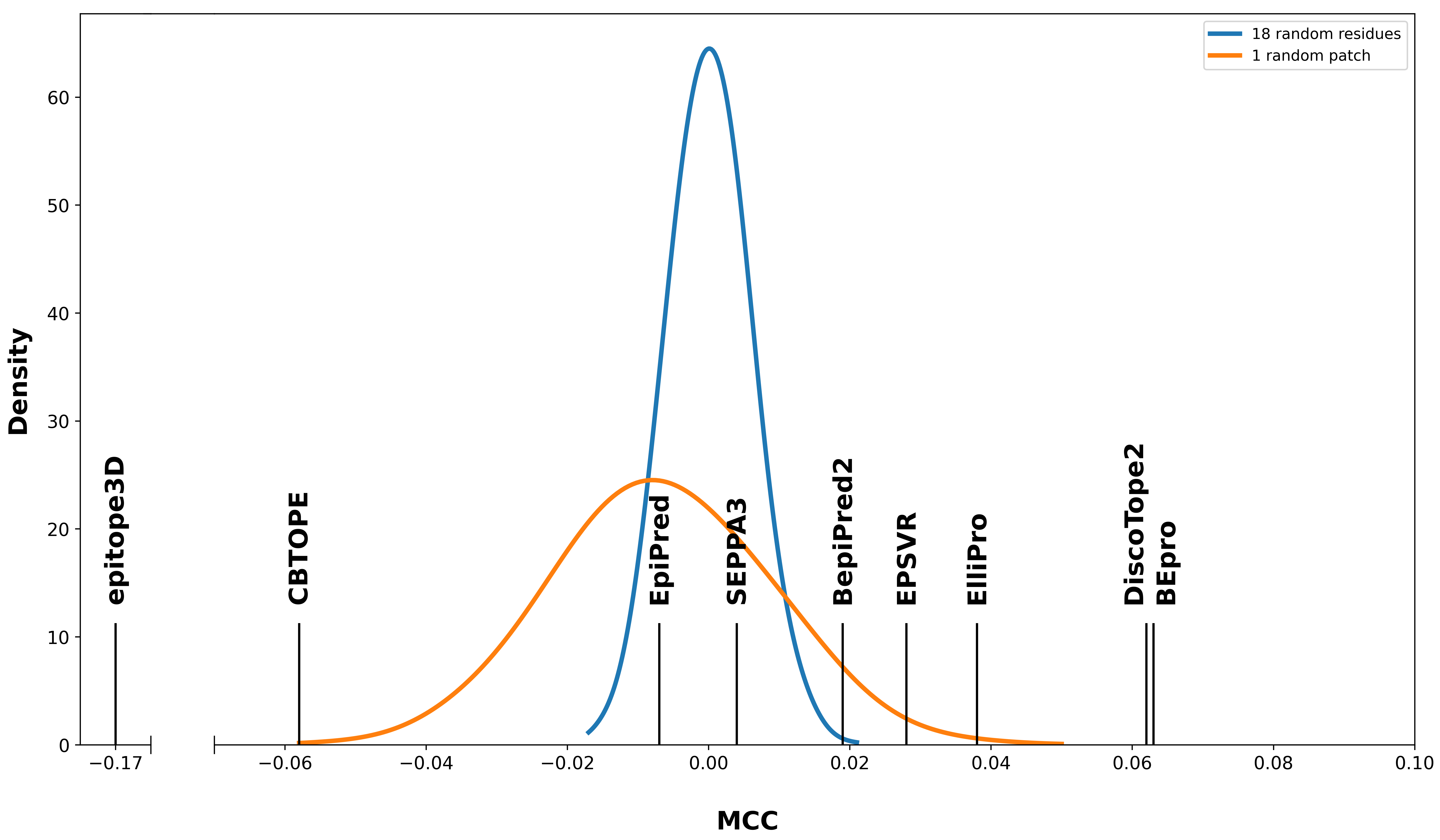}
\caption{Matthews correlation coefficient (MCC) of the benchmarked conformational B-cell epitope prediction methods, along with the bootstrap distributions obtained from a random procedure that generates 18 random surface residues (blue) or 1 random patch of 18 nearby surface residues (orange)  on each structure of the $\mathcal{E}_{Ag}^{rep}$ dataset, repeated 2,000 times.}
\label{fig:MCC_distribution}
\end{figure*}

To determine whether these results are statistically significantly better than random, we benchmarked the methods against a procedure which randomly predicts $N_r$ surface residues as epitopes. We tested two  strategies for the value of $N_r$: the first considers $N_r$ equal to the average size of an epitope, and the second one uses a dynamic $N_r$ that matches the number of epitope residues predicted by each method for each antigen (see Methods for details). As shown in \textcolor{black}{Table S2a}, when benchmarked against these strategies, only Bepipred2, DiscoTope2, ElliPro, EPSVR and BEpro are significantly better than our random procedure across all four metrics. SEPPA3 and EpiPred are significantly better only for some metrics, while CBTOPE and epitope3D are no better than random across all metrics.

We also benchmarked the methods against a random procedure that predicts patches of surface residues as epitopes instead of random residues scattered across the antigen surface (see Methods). This time only DiscoTope2, EPSVR and BEpro are significantly better than our random patches procedure across all metrics (\textcolor{black}{Table S2a}). BepiPred2, SEPPA3, ElliPro and EpiPred are significantly better for some of the metrics, while CBTOPE and epitope3D are no better than random patches across all metrics. Note that the reason behind the differences between the residue- and patch-based bootstrap distributions comes from the fact that the patch-based random procedure has a higher standard deviation than its residue-based counterpart (see \autoref{fig:MCC_distribution}). Indeed, predicting patches of residues leads to a higher possibility of predicting either many correct or incorrect residues than when predicting randomly scattered residues, given true epitopes are themselves patches of nearby residues.

One of the reasons that could explain why the majority of the methods did not perform better than random is the presence of false negative annotations in the dataset, corresponding to epitopes not yet identified. In order to improve the confidence in the epitope/non-epitope annotation of surface residues, we repeated the above benchmarking on the subset of $\mathcal{E}_{Ag}^{rep}$ consisting of antigens bound by at least 5 epitopes, noted  $\mathcal{E}_{Ag}^{rep(5)}$. EpiPred was excluded from this analysis given it is not affected by the issue of erroneous false negatives. The results are reported in \textcolor{black}{Figure S3} and \textcolor{black}{Table S2b}, which shows similar results than the previous evaluation on the full benchmark set, with BepiPred2, DiscoTope2 and BEpro performing better than random across all metrics.

In conclusion, all the methods showed very poor performances in absolute terms, and only two methods, namely DiscoTope2 and BEpro,  achieved better than random performances across all metrics and benchmarks. 

\subsection{Consensus predictions}

Often, conformational B-cell epitopes are predicted using a combination of several  methods and a consensus scheme whereby a residue is considered as an epitope if at least $ M$ methods predict it as such (see  \cite{devi2020exploring, olotu2021immunoinformatics, lon2020prediction, khare2021conformational} for recent examples). We therefore tested whether combining the predictions of all the generic epitope prediction methods gave better results than each one individually.

We first removed the structures that were in any of the selected methods' training datasets, resulting in a dataset of 65 structures on which the consensus predictions were evaluated. We  predicted a residue as an epitope if at least $M$ of the selected methods agreed. This consensus strategy gave the highest results for $M=4$, resulting in a ROC-AUC = 0.56, BAC = 0.56, MCC = 0.10 and PR-AUC = 0.34, which is slightly better than any individual predictor. Nonetheless, one should keep in mind that this result was obtained through optimization of the $M$ value in direct validation and is thus probably a bit overestimated.

In summary, even though consensus prediction schemes might be slightly better than single-method approaches or randomly predicted residues, they do not overcome the fundamental issue of the under-performance of B-cell epitope prediction methods.

\subsection{\textcolor{black}{Epitope} immunodominance}

The question of whether antibodies can be raised against any part of any antigen's surface has currently no definitive answer, but the poor performance of all the methods evaluated in the previous sections suggest a positive answer to this question. Even if the entire surface of any antigen can be bound by antibodies, some regions are undoubtedly targeted much more often than others by the immune system and more easily trigger the immune response. This phenomenon, known as epitope immunodominance \cite{kaur2009identification, ito2003immunodominance}, is important, for example when designing epitope-based vaccines that attempt to generate an immune response towards subdominant but functionally conserved sites where escaping mutations are less likely to occur \cite{paules2017pathway, angeletti2018possible, zost2019immunodominance}. Knowledge of the immunodominant and subdominant epitopes of an antigen can therefore be of great value.


One can reasonably expect that the scores attributed to each surface residue by conformational B-cell epitope predictors are correlated, at least to some extent, with residue immunodominance. To test this hypothesis, we estimated the immunodominance $\mathcal{I}$ of a given residue as the number of times it appears in an epitope; for this purpose, we restricted ourselves to the $\mathcal{E}_{Ag}^{rep(5)}$ dataset which only contains antigen structures that are bound by at least 5 different antibodies. As for the benchmarking, antigen structures that were part of the training dataset of a given method were removed. 
$\mathcal{I}$ was min-max scaled on a per-antigen basis to adjust for differences in number of epitopes per antigen structure. 

We computed the Spearman correlation between $\mathcal{I}$ and the predicted scores of each method, with the exception of EpiPred given immunodominance is irrelevant for antibody-specific methods.

\begin{table}[]
\begin{center}
\begin{tabular}{l c c}
 \hline
 Method & $r_{\mathcal{I}-{score}}$ &  $N_{residues}$\\
 \hline 
 BepiPred2 & \textbf{0.20}\textbf{*} &  3491 \\
  CBTOPE &	0.08\textbf{*}  &	 	5799 \\
 SEPPA3 & 0.15\textbf{*}  & 3107 \\
 DiscoTope2 & 0.15\textbf{*}  & 6287 \\
 ElliPro &  0.04  & 6910\\
 EPSVR &  0.09\textbf{*}  & 6490 \\
 BEpro & 0.14\textbf{*}  & 6287 \\
 epitope3D	& -0.03 & 4795\\
 \hline
\end{tabular}
\end{center}
\vspace{0.25cm}

\caption{
Spearman correlation coefficient $r_{\mathcal{I}-{score}}$ between the estimated immunodominance $\mathcal{I}$ and the per-residue epitope score outputted by each method on the $\mathcal{E}_{Ag}^{rep(5)}$ set. $N_{residues}$ corresponds to the number of residues on which the correlation was calculated. The highest correlation value is in bold. Due to the large sample size, we considered correlations as statistically significant if their $p$-value is $\le 0.001$, which are labeled with an asterisk.}
\label{table:benchmark_score_immunodominance_correlation}
\end{table}

As shown in \autoref{table:benchmark_score_immunodominance_correlation},  six out of the eight evaluated methods have a statistically significant Spearman correlation and are therefore better than random. However, the correlation values are very low, the highest  being 0.20 for BepiPred2; note that this is a sequence-based predictor. These results are in accordance with the low prediction scores observed in the previous subsection and indicate that the scores of the methods are not accurate enough to be used to deduce which epitopes are immunodominant. 

\textcolor{black}{Note that immunodominance is a highly complex phenomenon and that the $\mathcal{I}$ value that we used to estimate it, though intuitive, is clearly an approximation. Indeed, our dataset is biased towards  most (e.g., clinically) promising antibodies, and moreover contains highly engineered antibodies which do not necessarily reflect the preferences of the immune system. In addition,  $\mathcal{I}$ does not account for natural biases of the immune system such as antigenic imprinting \cite{vatti2017original} or original antigenic suppression \cite{angeletti2017defining}, where the immune system  preferentially uses or avoids the immunological memory based on previous infections.}

\subsection{SARS-CoV-2 case study}

As a case study, we evaluated the B-cell epitope predictors on the spike protein of the SARS-CoV-2 virus. This protein enables cell invasion through binding to the host's ACE2 receptor \cite{hoffmann2020sars}, and its 
receptor binding domain has been shown to be the preferential target of the host's immune response \cite{post2020antibody}. The SARS-CoV-2 spike protein is included neither in our $\mathcal{E}_{Ag}$ benchmark dataset nor in the training sets of the methods, and therefore constitutes an independent test case. \textcolor{black}{Note that another series of epitope prediction tools applied to SARS-CoV-2 has been reviewed in \cite{bukhari2022machine}.}

In order to evaluate the ability of the benchmarked methods to predict the known epitopes in the spike protein, we gathered 83 spike protein-antibody complexes resolved by X-ray crystallography or cryo-electron microscopy and deposited in the PDB \cite{berman2000protein} (see \cite{cia2022analysis,cia2022spikepro} for the list of PDB ids). We extracted 83 epitopes from these complexes; 75 of them are localized on the receptor binding domain (RBD) of the spike protein while the remaining 8 target its N-terminal domain (NTD). We evaluated the methods by mapping the 83 epitopes on the PDB structure 6VYB, which is a complete trimeric spike protein structure with one chain in open conformation \cite{walls2020structure}. Note that we do not have the results for all the predictors evaluated in the previous subsection as some of them failed to run on the large  protein trimer or their webserver was  down.

\begin{table}[h!]
\centering
\begin{tabular}{c c c c}
    \hline
         Method & MCC & ROC-AUC & $r_{\mathcal{I}-{score}}$ \\
    \hline
      BepiPred2 &  0.19 & 0.64\textbf{*} & 0.30\textbf{*} \\
         CBTOPE &  0.02 & 0.51 & 0.17\textbf{*} \\
     DiscoTope2 &  \textbf{0.27}\textbf{*} & 0.60 & 0.24\textbf{*}  \\
        ElliPro &  0.17 & 0.66\textbf{*} & 0.35\textbf{*}  \\
          EPSVR &  0.22 & \textbf{0.75}\textbf{*} & \textbf{0.45}\textbf{*}  \\
      epitope3D &  -0.20 & 0.38 & 0.039 \\
      \hline
\end{tabular}
\vspace{0.25cm}
\caption{MCC, ROC-AUC and Spearman correlation coefficient $r_{\mathcal{I}-{score}}$ between immunodominance  $\mathcal{I}$ and prediction scores of each method on the SARS-CoV-2  spike protein trimer. Statistically significant results ($p$-value $\le$ 0.001) are labeled with an asterisk. The highest value for each metric is in bold.}
\label{table:SARS_MCC_PPV_results}
\end{table}

\begin{figure*}[h!]
\centering
\includegraphics[width=\textwidth]{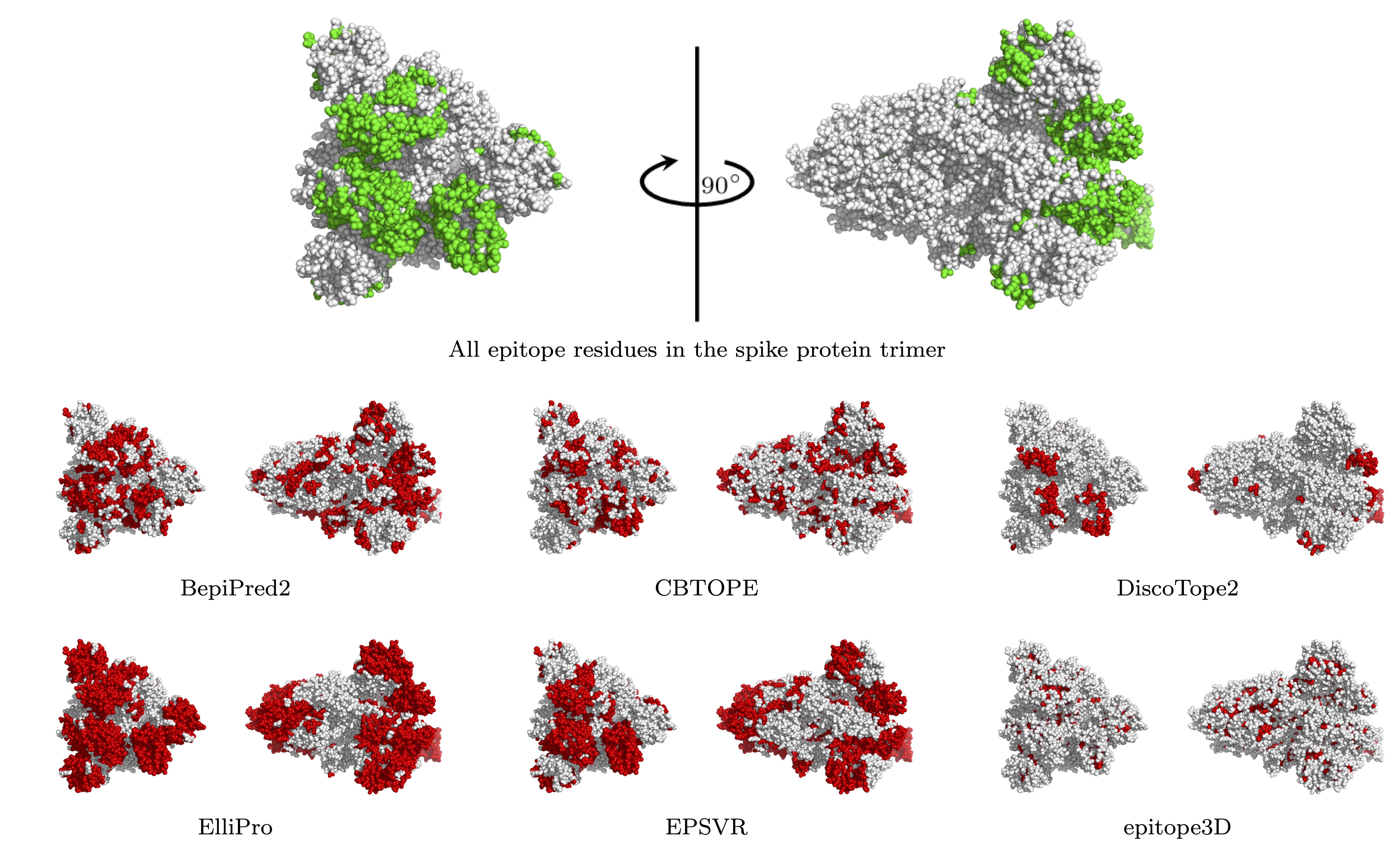}
\caption{Visual representation of all the true epitope residues of the SARS-CoV-2 spike protein trimer (green) and the residues predicted by each of the evaluated conformational B-cell epitope prediction methods (red). 
All the figures were generated with PyMOL \cite{delano2002pymol}}
\label{fig:SARS_method_predictions}
\end{figure*}

The prediction scores  of the methods are given in \autoref{table:SARS_MCC_PPV_results} and the localization of the predicted and real epitopes in the 3D structure of the spike protein trimer are shown in \autoref{fig:SARS_method_predictions}. Some of the predictors have relatively good scores, especially when compared with the performances on the large benchmark dataset analyzed in the previous subsection. EPSVR obtained the best results, reaching a ROC-AUC of 0.75 and a score-immunodominance correlation of 0.45. This can clearly be seen in \autoref{fig:SARS_method_predictions} where EPSVR predicts very well a large portion of the RBD and NTD epitope residues. 
The worse performing method is again epitope3D which, as previously observed, is biased towards non-epitope core residues. The remaining methods did not perform too well, as they either overpredicted (ElliPro), underpredicted (DiscoTope2) or made predictions all over the surface (BepiPred2 and CBTOPE).

\subsection{Per-antigen performance analysis}
The fact that the overall SARS-CoV-2 results are better than those on the large benchmark is interesting and prompted us to dig further into our results. We analyzed the methods' performances on a per-antigen basis by  plotting the distribution of MCC and Spearman correlation coefficients  of each method; they are  shown in \textcolor{black}{Figures S4 and S5}. Both figures show that all the methods have very variable performances according to the  antigen, in other words, they have a high standard deviation. Indeed, some entries are well predicted with high scores, while others are  completely wrongly predicted. 

It could be interesting to further analyze such results to understand whether the well predicted antigens are due to randomness, to  biases towards the method's training dataset,  or to  truly learned features that distinguish epitope and non-epitope residues. Understanding this could boost the development of improved predictors and help  to better understand their reliability.   

Note that, despite the fact that predictions on the SARS-CoV-2 spike protein show  better results on the average,  B-cell epitope predictors did not contribute much to antibody development during the pandemic. Indeed, experimental characterization of antibodies starting from plasma of infected COVID-19 patients followed by 3D cryo-electron microscopy has been the method of choice to design therapeutic monoclonal antibodies \cite{taylor2021neutralizing}. 

\section{Conclusion}

Many \textit{in-silico} tools to predict conformational B-cell epitopes using sequence- and/or structure-derived features have been published in the last 20 years. 
They  have all compared themselves to each other and almost systematically claim to outperform   all the other methods. However, no independent benchmarking has been published in recent years \cite{yao2013conformational}.
In this paper we have carefully assessed nine of the most popular and recent prediction methods available through a webserver, including eight generic predictors and an antibody-specific predictor, on a large and well-curated set of antigen-antibody structures. 

Our benchmarking results show that the overall performance of the methods is very poor, and that many of them do not perform significantly better than randomly predicted patches of residues. Indeed, only 2 out of the 9 evaluated methods perform significantly better than random both on the benchmark dataset and on the subset of antigens for which we have the highest confidence about epitope/non-epitope residue labels. \textcolor{black}{Note that some of the tested methods were trained over 10 years ago and  their performance might have been better if they had been retrained on an up-to-date training dataset.}

In addition, we evaluated the performance of consensus strategies that combine multiple predictors, which is a frequently used approach in B-cell epitope prediction. Our  results show that even this strategy is not much better than random predictions.

Regarding the evaluation methodology used in this benchmark, we combined both threshold-dependent (BAC and MCC) and threshold-independent (ROC-AUC and PR-AUC) metrics in order to capture the overall performance of the methods. They all point towards the same conclusion regarding the predictive performance of the methods, which is much lower than what we expected. 

Our benchmarking also provides hints about open questions and improvements that could be made to current prediction methods. First of all, our analysis highlights the pervasive issue of the incomplete knowledge of all the true epitopes of any given protein, especially in lesser studied ones.  This has a major impact on evaluation procedures given there is always the uncertainty that a non-epitope residue might be part of a yet unknown epitope. For that reason, we performed our benchmark analysis on both the full dataset of representative structures $\mathcal{E}_{Ag}^{rep}$ and on the subset of antigens with at least 5 mapped epitopes, in order to assess if there is a difference in the methods' performance when evaluated on antigen structures for which we have a higher confidence in the labels of their surface residues, but our results showed no significant difference between these two benchmarks.
 
Related to this, the question arises of whether an antibody can be obtained against any surface region or if only certain parts with favourable structural \textcolor{black}{and} physico-chemical features are valid candidates. \textcolor{black}{In the former case, the strong immunodominance observed in nature could originate from biases specific to the naive immune system of each species.}
Although our benchmarking analysis cannot answer this question with certainty, the systematically low performance of all tested methods suggests that it is indeed the case. However, some regions that are more easily recognized by antibodies, known as immunodominant epitopes, have been shown to elicit a stronger immune response \cite{frank2020immunology} and it should thus be possible to identify their characteristic features. A subsidiary question is whether natural and engineered antibodies are different in that respect.

Another interesting observation from our results is  that each method predicts some proteins quite well, with  high prediction scores, and others very  poorly. The analysis of these outliers could be helpful to understand the advantages and limitations of each method and help design better performing methods. It must be noted that the predictors do not agree in general: a protein that is well predicted by one method is usually not well predicted by the others.

Regarding the features used by the different methods, we found that the large majority of predictors overlook some important features whose consideration would certainly boost their performances:  

\begin{itemize}
 \item Glycosylated antigen regions usually cannot be recognized by antibodies due their shielding effect \cite{chang2019glycosylation, casalino2020beyond,wintjens2020impact}. The annotation or prediction of glycosylated regions should thus be included in the predictors to boost their performances.
 
 \item Antigens can undergo conformational changes where some regions get masked and become unavailable for antibody binding. The spike protein trimer of SARS-CoV-2 is an example of this: it occurs in  open and closed conformations characterized by differences in solvent accessibility and  epitopes  \cite{cia2022analysis}. 
 
 \item  Oligomerization  properties of antigens are also important for determining their immunogenicity. Indeed, oligomers hide regions from the solvent which are then no longer accessible to antibodies and, at the same time, lead to inter-chain surface regions that can be targeted by antibodies. Prediction methods often do not take these properties into  account.
 
 \item It would be beneficial for webservers to give users the ability to provide additional information about the target protein, such as residues that cannot be bound and should therefore be ignored by the predictor.
 
 \item Another interesting improvement would be the possibility for the user to provide antibody sequences for which he wishes to identify  potential epitopes, as this has been suggested to improve prediction performances \cite{jespersen2019antibody}.
 
\item Finally, the application of recent advances in Natural Language Processing (NLP) could enable the development of novel methods \cite{clifford2022bepipred,park2022epibertope} that  help advance the field. 
\end{itemize} 

We hope this work will be of use for future research in B-cell epitope prediction and  help solve some of the critical issues. It is important to set up additional independent benchmarks as well as blind prediction experiments 
as they would contribute to a better understanding of the biases and limitations of epitope prediction methods and advance the field.

\section{Competing interests}
There is NO Competing Interest.

\section{Acknowledgments}
We thank Basile Stomatopoulos for interesting discussions on the subject. We acknowledge  financial support from the Fund for Scientific Research (FNRS) through a COVID-PER project and a PDR project. M.R. and F.P.  are FNRS research director and postdoctoral researcher, respectively, and G.C. benefits from a FNRS-FRIA PhD grant.

\section{Author contributions statement}
Conceptualization, F.P. and M.R.; G.C. conducted the experiment(s). All author analysed the results, wrote and reviewed the manuscript, and agreed to the published version of the manuscript.

\bibliographystyle{unsrt}
\bibliography{references}

\end{document}